# Planetary Nebulae in the dwarf galaxy NGC 6822: Detection of new candidates ⋆, ⋆⋆

P. Leisy[1,2], R.L.M. Corradi[1], L. Magrini[3], R. Greimel[1], A. Mampaso[2], and M. Dennefeld[4]

[1] Isaac Newton Group Telescope, Sea Level Office, Apartado de Correos 321, E-38700 Santa Cruz de La Palma, SPAIN
[2] Instituto de Astrofísica de Canarias, c. Via Láctea s/n, E-38200, La Laguna, Tenerife, SPAIN
[3] Dipartimento di Astronomia e Scienza dello Spazio, Universitá di Firenze, L.go E. Fermi 2, I-50125 Firenze, ITALY
[4] Institut d'Astrophysique de Paris, 98bis, Boulevard Arago, F-75014 Paris, FRANCE

**Abstract.** We have discovered 13 new candidate Planetary Nebulae (PNe) in the dwarf irregular galaxy NGC 6822, increasing its total number to 17 objects. Our sample of candidate PNe is complete down to 3.5 mag below the brightest PN.

## 1. Introduction

Planetary Nebulae (PNe) are prime targets to study the chemical evolution of nearby galaxies, as they are excellent tracers of the metallicity of stellar populations over a wide range of ages and in all types of galactic and intergalactic components.

We are carrying out an imaging programme (the Local Group Census) to search for PNe in the Local Group (LG) galaxies both in the Northern and Southern hemispheres. One spiral (M 33), and several irregular and spheroidal galaxies have been analyzed (Magrini et al. 2000, Magrini et al. 2002, Magrini et al. 2003, Corradi et al. 2004), significantly increasing the population of known PNe in the LG.

NGC 6822 is a barred dwarf galaxy of type Ir IV-V with a total luminosity of $M_V = -15.96$ and a mass of $1.9\,10^9\,M_\odot$ (van den Bergh 2000). It is brighter, larger, and closer than any other dwarf irregular in the LG (Mateo 1998), allowing deeper studies of the properties of individual stars and emission-line objects. The optical diameter is about $2.9 \times 2.9$ kpc (Hodge et al. 1991) with a small optical bar of $0.9 \times 0.9$ kpc. However, a study of carbon stars reveals that the stellar population of NGC 6822 extends out to more than $4 \times 6$ kpc from the center (more than 5 scale lengths, Letarte et al. 2002). Moreover, a radio survey shows a huge H I halo extending to $6.1 \times 12.9$ kpc (de Blok & Walter 2000).

NGC 6822 is gas rich with clear evidence of recent star formation. The star formation rate was roughly continuous and low during the last 1 Gyr (Marconi et al. 1995). This has been confirmed by Gallart et al. (1996) and Wyder (2001), who concluded that NGC 6822 continuously formed stars during its early evolution (12-15 Gyr ago), with possible modest enhancements 400 and 100-200 Myr ago.

Killen & Dufour (1982) classified 36 stellar emission objects and 31 diffuse extended objects: 8 PNe (KD S29 is also classified as an H II region), 14 small/compact H II regions, and 14 H$\alpha$ emission line stars. Hodge et al. (1988) identify 157 H II regions, while Collier & Hodge (1994) and Collier et al. (1995) studied their luminosity and morphology.

One SNR has been found by D'Odorico et al. (1980) and Killen & Dufour (1982, object KD D20), apparently associated with a variable X ray source. In their recompilation Ford & Peng (2002) reported 7 candidate PNe of which only limited spectroscopic information is available in the literature: one spectrum (KD S33) by Dufour & Talent (1980), and another one (KD S16) by Richer & McCall (1995). Both found an abundance log(O/H)+12=8.1±0.1, equal within errors to the value of 8.14 derived by Skillman et al. (1989) for H II regions.

More data exist for H II regions (Lequeux et al. 1980, Hodge et al. 1989, Chandar et al. 2000, Hidalgo-Gámez et al. 2001, O'Dell et al. 1999), but a comprehensive, homogeneous and accurate H II abundance database is still missing. New abundance determinations for both PNe and H II regions are clearly needed in order to allow a proper comparison with the recent determinations for individual B (Muschielok et al. 1999) or A (Venn et al. 2003) supergiant stars.

Our observations and data reduction are described in Sect. 2. The object selection is presented in Sect. 3 and the final catalog is given in Sect. 4. The results are discussed in Sect. 5.

## 2. Observations

### 2.1. Imaging

In a number of observations, both from La Silla (ESO) and La Palma (ORM), we searched for candidate PNe by means of narrow-band imaging using prime-focus wide-field imagers, which largely reduced the amount of observing time needed to cover the relatively large field of NGC 6822.

**Table 1.** ESO_WFI observations in July-August 1999.

| Filter | Exp.(s) | Seeing | Airmass |
|---|---|---|---|
| 30/07/1999 | | | |
| B | 3x 60 | 1.0" | 1.16 |
| [O III] | 5x 540 | 1.0" | 1.04 |
| Hα | 5x 360 | 0.9" | 1.12 |
| [S II] | 5x 720 | 0.9" | 1.06 |
| Cont | 5x 360 | 1.0" | 1.09 |
| 31/07/1999 | | | |
| B | 3x 120 | 1.0" | 1.77 |
| [O III] | 5x 540 | 1.2" | 1.08 |
| Hα | 5x 360 | 1.0" | 1.66 |
| [S II] | 5x 720 | 0.9" | 1.41 |
| Cont | 5x 360 | 1.1" | 1.16 |

**Table 2.** INT_WFC observations

| Name | Date | Exp.(s) | Seeing | Airmass |
|---|---|---|---|---|
| North West | $\alpha = 19h44m41.4$  $\delta = -14°36'11.0''$ | | | |
| [O III] Str. $y$ | 24/04/2001 | 600-1200 | 1.3-1.6 | 1.71-1.40 |
| Hα g r | 18/09/2001 | 400-1200 | 1.3-1.7 | 1.37-1.70 |
| North East | $\alpha = 19h46m08.2$  $\delta = -14°34'11.0''$ | | | |
| [O III] Str. $y$ Hα | 19/09/2001 | 600-1200 | 1.3-1.8 | 1.37-1.80 |
| [O III] Str. $y$ | 25/09/2001 | 600-1200 | 1.6-2.4 | 1.37-1.65 |
| Hα r g | 25/09/2002 | 300-400-1200 | 1.0-1.2 | 1.38-1.74 |
| [O III] Str. $y$ Hα r | 03/10/2002 | 200-300-1200 | 1.1-1.8 | 1.37-1.83 |
| Hα r g | 04/10/2002 | 200-300-1200 | 1.2-1.7 | 1.40-1.73 |
| [O III] Str. $y$ | 01/05/2003 | 600-1200 | 1.5-1.6 | 1.63-1.46 |
| South West | $\alpha = 19h44m41.4$  $\delta = -14°57'11.0''$ | | | |
| Hα r [O III] Str. $y$ | 03/06/2002 | 400-600-1200 | 1.5-1.9 | 1.73-1.38 |
| Hα g | 25/09/2002 | 300-400-1200 | 1.0-1.2 | 1.38-1.74 |
| Hα g | 04/10/2002 | 200-300 | 1.4-1.6 | 1.40-1.73 |
| [O III] | 05/10/2002 | 1200 | 1.5-1.8 | 1.49-1.65 |
| r | 08/10/2002 | 300-1200 | 1.6-2.1 | 1.41-2.15 |
| South East | $\alpha = 19h46m04.1$  $\delta = -14°56'11.0''$ | | | |
| Hα r [O III] Str. $y$ | 02/06/2002 | 300-600-1200 | 1.0-1.4 | 1.64-1.38 |
| g | 25/09/2002 | 300-400-1200 | 1.0-1.2 | 1.38-1.74 |
| [O III] Str. $y$ | 27/09/2002 | 600-1200 | 1.1-1.8 | 1.38-1.69 |
| [O III] g | 04/10/2002 | 200-300 | 1.2-1.7 | 1.40-1.73 |
| r g | 30/04/2003 | 600 | 1.4-1.6 | 1.59-1.42 |
| North West offset | $\alpha = 19h43m00.0$  $\delta = -14°25'00.0''$ | | | |
| [O III] Str. $y$ | 02/10/2003 | 300-1200 | 1.4 | 1.37 |
| g r | 02/05/2003 | 600 | 1.1-1.5 | 1.51-1.39 |

Table 1 presents the logs of the imaging observations obtained with the Wide Field Imager (WFI) at the ESO/MPA 2.2m telescope at La Silla. The center of the observed field was at $\alpha = 19h45m00.0$  $\delta = -14°47'45.0''$. The filters used have the following central wavelength and full width at half maximum (FWHM): 658/7 nm (Hα), 503/8 nm ([O III]), 676/8 nm ([S II]), and 604/21 nm ("continuum" filter especially suited for Hα). In each filter, five dithered exposures were taken in order to cover the CCD gaps between the 8 CCDs and reconstruct a full frame of 34′ x 34′. All nights were photometric, and several standards stars (LTT 4816, LTT 6248, LTT 7379, LTT 7987) were observed, allowing accurate flux calibration.

The observations at La Palma with the 2.5m INT telescope and its Wide Field Camera (Table 2) cover a larger area. Five overlapping fields were observed, providing a total field of view of 68×56 arcmin$^2$. The filters used were Hα (656.8/9.5nm), r (624/135nm), [O III] (500.8/10nm), and Str. $y$ (550.5/24nm). The data were also flux calibrated using standard stars in photometric nights. More details on the camera, a mosaic of four $2 \times 4$ K EEV CCDs, and the filters used, can be found in previous papers (e.g. Magrini et al. 2000).

In both telescopes, the Hα filter also includes the [N II] emission-line doublet at 654.8 and 658.3nm. The limiting magnitudes of the searches with both telescopes are similar and reach about 24 in [O III] (see Table 3).

### 2.2. Data reduction

The WFI observations have been reduced mostly using the MSCRED IRAF [1] package (V 0.1) with the aid of a special dedicated astrometry tool developed by Luca Rizzi in Padova (Held et al. 2001). This tool mainly uses the MSCRED commands, but in a much more automatic way, and allows us to obtain precise astrometry. The following steps were used for the data reduction:

1. FITS headers manipulation (*esohdr* task);
2. Trimming, bias substraction, and flat-field division;
3. Re-centering of frames (Continuum filter as reference);
4. Finding the astrometric solution;
5. Creating a full image frame ($8 \times 8$ K);
6. Matching the photometric parameters to scale each image against a reference image;
7. Stacking all the 5 dithered images per filter into the single image;
8. Extracting the sources with *SExtractor* (Bertin & Arnouts 1996);
9. Flux calibration.

A more detailed description of each individual step will be given in a forthcoming paper (Leisy et al., *in preparation*). The INT/WFC images were reduced through the ING WFC data-reduction pipeline (Irwin & Lewis 2001), and then analyzed as described in Magrini et al. (2000) and Corradi et al. (2004).

On the ESO WFI field, we determined an astrometric solution using the USNO-A2.0 catalog (Monet et al. 1998) with residuals smaller than 0.2″. A totally independent solution with about the same precision was obtained for the ING WFC image.

---

[1] IRAF is distributed by the National Optical Astronomy Observatories, which are operated by the Association of Universities for Research in Astronomy, Inc., under cooperative agreement with the National Science Foundation

## 3. Selection of candidate PNe

We first produced an image for each filter, and then continuum-subtracted images in [O III], H$\alpha$, and [S II]. These images allow us to select candidate PNe.

For the WFI images, a properly scaled image subtraction of the nearby continuum filter provides good removal of background stars for H$\alpha$. Unfortunately, the blue (B), or the red continuum filters used as a continuum for [O III] does not give equally good results. Thus [O III] and H$\alpha$ continuum-subtracted images were only used for a first visual identification of candidate PNe. Fluxes for the candidates were measured for each filter using *SExtractor*.

We selected the candidate PNe using the following criteria:

- To be detected in both the [O III] and H$\alpha$ filters, but not in the continuum at 604 nm, or at least more than 15 times fainter than in H$\alpha$ [2] (the maximum ratio for H II regions in Table 3, see also Sect. 4.1);
- To be point-like. At 500 kpc, 1″ corresponds to 2.4 pc, much bigger than the typical size of the bright inner body of a PN;
- The flux ratio $\frac{[S\,II]}{H_\alpha} \lesssim 0.3$ to exclude probable SNR.

In the WFC images, continuum-subtraction was fairly good and, as described by e.g. Magrini et al. (2000), candidate PNe were selected directly from the continuum-subtracted images as spatially unresolved sources that emit both in H$\alpha$ and [O III]. An additional check of the selection was done using color-color diagrams, as in Corradi et al. (2004).

As in our previous papers (Magrini et al. 2000), we do not use the excitation parameter R=[O III]/H$\alpha$ to discriminate between PNe and compact H II regions, as the fraction of genuine PNe with a low ($R < 1$) ratio is not negligible: they represent 25% of the global sample of PNe in the Milky Way, and ~20-35% in the LMC, SMC, M33, and NGC 6822.

A few objects of uncertain classification have also been found. Seven of them, after careful study of the H$\alpha$ and [O III] images, were rejected as possible PNe and classified as compact H II regions because they were found to be extended. Three other objects (KD_S23, KD_S26, and KD_S29 in Killen & Dufour 1982) previously listed as candidate PNe were rejected on the basis of no evident line emission and a relatively strong continuum emission (fluxes of 242, 1385, and 128 $10^{-16}$ erg $cm^{-2}$ $s^{-1}$, respectively). They are therefore stars. In color-color diagrams they also fall on top of the main stellar locus, further justifying their classification as stars.

## 4. List of candidate PNe

Table 3 presents the 17 candidate PNe in NGC 6822 selected in our images, as well as 9 objects that were finally classified as compact H II regions.

In Column 1 of the Table, we assign an ID number to each candidate, according to our initial selection, preceded by "pn" or "H II" depending on whether we classify it as a PN or a compact H II region. Objects common to the catalog of Killen & Dufour (1982) are also indicated as numbers with the "KD_S" prefix. Col. 2 and 3 list the objects coordinates. In Col. 4 to 7 we list the undereddened fluxes derived from our WFI images, and in col. 8 and 9 line ratios which help in determining if a candidate is a PN, H II region or SNR. In Col. 10 we present the [O III] magnitude calculated from the [O III] fluxes according to the usual formula $m_{[O\,III]}$ = -2.5 $\log(F_{[O\,III]})$-13.74 (Jacoby 1989). For the objects in common, our fluxes are in agreement with those measured by other authors like Dufour & Talent (1980) or Hodge et al. (1988). Finally in Col. 11 some comments are made for the extended objects.

---

[2] In very deep surveys, the brightest PNe clearly show a visible continuum

Note that all the 17 candidates PNe but pn-003 and pn-019 were already detected and their velocities measured by Jacoby (unpublished data) as emitting line objects in [O III]$_{5007}$ and [O III]$_{4959}$ (only the brightest line for pn-005, pn-009, and pn-016). All the sources are found in both the WFI and WFC images, and no additional candidate was found in the outermost regions of the larger area of sky covered by the WFC observations.

### 4.1. Comments on individual objects

Killen & Dufour (1982) classified 8 objects as PNe (KD_S29 as both a PN or an H II region), but we can only confirm four of them (KD_S14, KD_S16, KD_S30, and KD_S33). KD_S10 is extended and is definitively not a PN. As already mentioned before, KD_S23, KD_S26, and KD_S29 are stars. KD_S28 is not an emission line star, but an H II region.

Two bright objects (KD_S2 and KD_S5) of Killen & Dufour (1982) were identified as H II regions by Kinman et al. (1979, named respectively $K_\alpha$ and $K_\beta$ in Table 3). In our images these objects are resolved, confirming them as compact H II regions. Five of the faint compact H II regions (H II-20 to H II-24) could be SNR, given their large [S II]/H$\alpha$ ratio.

Note that one can also use the ratio H$\alpha$/Cont to test the robustness of our identification. Compact H II regions have a ratio which ranges from 2 to 15, while all our candidate PNe have much larger ratios, up to 200. Indeed most have been confirmed as PNe by follow-up spectroscopy (Leisy et al. *in preparation*).

### 4.2. Finding charts

In addition to accurate coordinates, to help future observations, we present finding charts for each object in Table 3 as 1.5′×1.5′ grey-scale plot of the H$\alpha$ image. All objects are centered and indicated on the finding charts. The 17 candidate PNe are shown in Fig. A.1, and the H II regions or possible SNRs in Fig. A.2.

## 5. Discussion

### 5.1. Distribution of candidate PNe within the galaxy

As shown in Fig. 1, most of the candidate PNe lie in the bar of NGC 6822. De Blok & Walter (2003) found that the young

**Table 3.** NGC 6822 list of candidate PNe and image fluxes

| name | $\alpha$ (J2000) | $\delta$ (J2000) | Fluxes in $10^{-16}.erg.cm^{-2}.s^{-1}$ | | | | Ratio | | $m_{[O\,III]}$ | comments |
|---|---|---|---|---|---|---|---|---|---|---|
| | | | [O III] | H$\alpha$ | [S II] | Cont | [O III]/H$\alpha$ | [S II]/H$\alpha$ | | |
| pn-001 | 19 44 02.287 | -14 42 42.81 | 225.0 | 91.0 | <2.5 | <1.5 | 2.47 | <0.03 | 20.38±0.10 | |
| pn-002 | 19 44 49.099 | -14 42 58.59 | 76.0 | 49.5 | <1.0 | <1.5 | 1.53 | <0.03 | 21.55±0.15 | |
| pn-003 | 19 44 51.988 | -14 42 15.79 | 11.0 | 26.0 | <2.5 | <1.5 | 0.43 | <0.10 | 23.64±0.40 | |
| pn-004=KD_S30 | 19 45 01.568 | -14 41 36.79 | 372.0 | 165.0 | <1.0 | <1.5 | 2.25 | <0.01 | 19.83±0.05 | |
| pn-005 | 19 45 56.401 | -14 40 50.13 | 61.0 | 37.5 | <1.0 | <1.5 | 1.63 | <0.03 | 21.80±0.15 | |
| pn-006 | 19 46 02.228 | -14 43 41.61 | 32.5 | 47.0 | <1.0 | <1.5 | 0.69 | <0.03 | 22.47±0.20 | |
| pn-007=KD_S33 | 19 45 07.180 | -14 47 30.88 | 378.5 | 287.0 | <1.0 | <1.5 | 1.32 | <0.00 | 19.81±0.05 | |
| pn-009 | 19 45 06.483 | -14 48 39.36 | 12.5 | 12.5 | <1.0 | <1.5 | 1.00 | <0.10 | 23.46±0.40 | |
| pn-010 | 19 44 59.764 | -14 48 07.46 | 92.0 | 21.5 | <1.0 | <1.5 | 4.29 | <0.06 | 21.35±0.15 | |
| pn-011 | 19 44 51.825 | -14 48 16.35 | 6.5 | 7.5 | <2.5 | <1.5 | 0.88 | <1.54 | 23.49±0.40 | |
| pn-012 | 19 44 50.945 | -14 49 14.33 | 30.0 | 14.5 | <1.0 | <1.5 | 2.07 | <0.09 | 22.57±0.30 | |
| pn-013=KD_S16 | 19 44 49.538 | -14 48 04.12 | 203.0 | 96.5 | <1.0 | <1.5 | 2.10 | <0.01 | 20.49±0.10 | |
| pn-014=KD_S14 | 19 44 49.578 | -14 46 29.61 | 219.0 | 65.5 | <1.0 | <1.5 | 3.35 | <0.02 | 20.41±0.10 | |
| pn-015 | 19 44 31.253 | -14 47 13.52 | 23.0 | 29.0 | <1.0 | <1.5 | 0.78 | <0.04 | 22.86±0.40 | |
| pn-016 | 19 44 45.599 | -14 52 37.40 | 17.5 | 4.0 | <1.0 | <1.5 | 4.18 | <1.50 | 23.15±0.40 | |
| pn-017 | 19 44 44.263 | -14 50 46.83 | 30.5 | 19.5 | <1.0 | <1.5 | 1.56 | <0.06 | 22.54±0.40 | |
| pn-019 | 19 44 56.932 | -14 45 17.77 | 157.5 | 134.5 | <2.5 | <1.5 | 1.17 | <0.02 | 20.76±0.10 | |
| Extended objects | | | | | | | | | | |
| H II-008=KD_S28 | 19 44 57.152 | -14 47 50.58 | 1338.5 | 1014.5 | 207.0 | 76.0 | 1.32 | 0.20 | 18.44±0.03 | |
| H II-018=KD_S10 | 19 44 42.613 | -14 50 30.24 | 317.5 | 534.0 | 100.0 | 5.0 | 0.59 | 0.19 | 20.00±0.05 | |
| H II-020 | 19 44 54.497 | -14 42 42.41 | 56.0 | 182.0 | 78.0 | 58.0 | 0.31 | 0.43 | 21.89±0.15 | SNR? ext? |
| H II-021 | 19 44 56.711 | -14 42 05.92 | 12.0 | 58.0 | 20.5 | 11.5 | 0.21 | 0.36 | 23.54±0.40 | SNR? ext? |
| H II-022 | 19 44 52.279 | -14 42 48.31 | 10.0 | 51.5 | 22.5 | 12.0 | 0.19 | 0.43 | 23.76±0.40 | SNR? |
| H II-023 | 19 44 52.130 | -14 42 53.99 | 11.0 | 36.0 | 18.5 | 21.0 | 0.30 | 0.51 | 23.66±0.40 | SNR? |
| H II-024 | 19 44 58.514 | -14 44 45.90 | 26.0 | 81.0 | 61.0 | 33.0 | 0.32 | 0.75 | 22.72±0.40 | SNR? |
| H II-KD_S2=$K_\alpha$ | 19 44 30.905 | -14 48 25.28 | 5752.0 | 3754.0 | 449.0 | 377.0 | 1.53 | 0.12 | 16.86±0.03 | |
| H II-KD_S5=$K_\beta$ | 19 44 32.980 | -14 47 30.20 | 2599.5 | 2024.5 | 315.0 | 149.5 | 1.28 | 0.16 | 17.72±0.03 | |

blue stars are mostly located in the bar, and recently, Battinelli et al.( 2003) observed that the spatial distribution of massive stars matches the distribution of H II regions well. It would be very interesting to measure the abundances of the candidate PNe and the ages of their central stars, to determine if they originated from the old population of the bar, or inspead, from relatively young and massive progenitors

Notably, three PNe were discovered at large galactocentric distances (see Fig. 1), about 15′ (∼2.2 kpc) from the center of NGC 6822, which itself is surrounded by a large neutral hydrogen envelope with an inclined elliptical shape (Hutchmeier 1979). One is tempted to associate these 3 PNe with this large H I envelope. They could be the first stellar representatives, but at least the 2 eastern PNe seem not to be spatially associated. They could also be inter-galactic PNe, a hypothesis that requires confirmation by spectroscopy. Note that Magrini et al. (2003) found 4 candidate PNe clearly outside the optical diameter of another dwarf irregular galaxy, IC 10, which is also known to possess an extended H I halo.

### 5.2. PNe Number

The number of candidate PNe in each galaxy of the LG scales pretty well with the total luminosity of the galaxy (see Magrini et al. 2003 and the updated graph in the recent review of Corradi & Magrini 2004). The relation is usually normalized using the data for the SMC, the galaxy in which the observed PN population size is the most complete, covering a range of at least 6 mag (Jacoby & De Marco 2002, see also Jacoby 2004). The number of PNe found in NGC 6822 falls close to this relation, making us confident of having found all the brightest objects. Using the observed PNLF of the SMC (Jacoby & De Marco 2002) and scaling with the luminosity of NGC 6822, we calculate that about 50% of the total number of PNe have been found.

Following Ciardullo et al. (1987), we estimate the completeness limit of our survey when the signal to noise of the objects is greater than 10. This limit corresponds to a magnitude of $m_{[O\,III]}$ = 23.35. Therefore we reach 100% completeness over the brightest 3.5 magnitude range of the PNLF.

### 5.3. PNe Luminosity Function - distance

Since 1989, the [O III] PNLF has been widely used as a distance indicator (Ciardullo et al. 1989). In the case of NGC 6822 the number of PNe is small, and any distance determination through the PNLF must be made with caution (cf. Méndez et al. 1993, Magrini et al. 2003). Moreover this also results in a large uncertainty on the derived PNLF distance. The Eddington formula (Eddington 1913) was applied to each bin of our data in order to correct for the effects of a known probable observational error (see Ciardullo et al. 1989).

For this small sample of PNe, we can derive a crude distance modulus taking only the brightest PN (m-M=24.3) into account. In Fig. 2 the PNLF with the completeness limit is shown. We also plotted our best fit for these points.

An uncertain fit to the "universal" PNLF (Jacoby 1989) could only be obtained for the two first bins of 0.6 mag, that is. on 6 of the 17 PNe found. With this small sample we obtained a similar distance modulus value of 24.1, but with a large uncertainty (see for example the discussion in galaxy M 32 (Ciardullo et al. 1989)).

There is an interesting point to note in Fig. 2: The fit is complicated by a dip after the 2 first bins, which look similar (but much less statistically significant) to the dip, around 4 mag from the brightest PN, in the PNLF of the SMC observed by

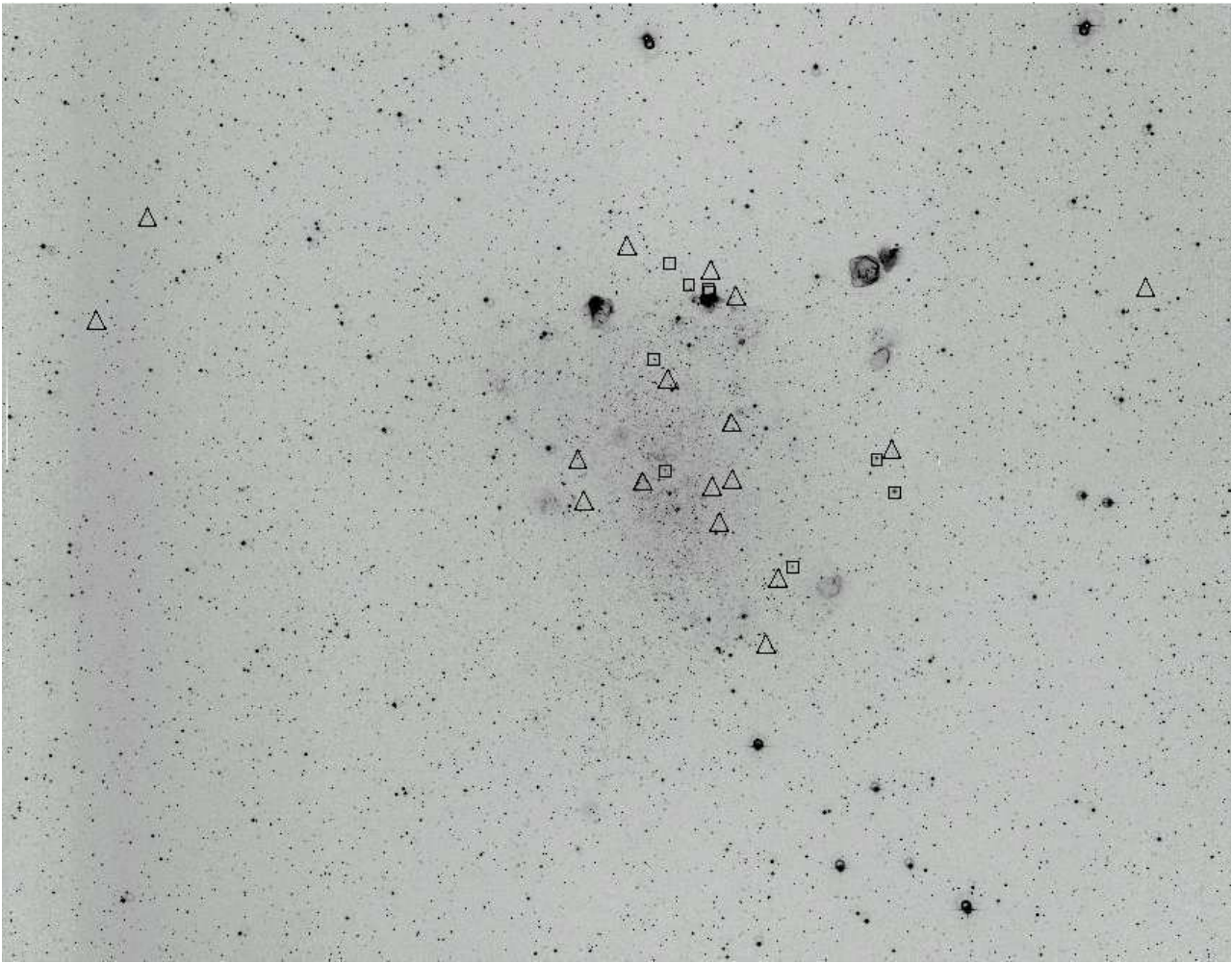

**Fig. 1.** Location of emission line objects in an H$\alpha$ image of NGC 6822: PNe (triangles), H II regions (squares)

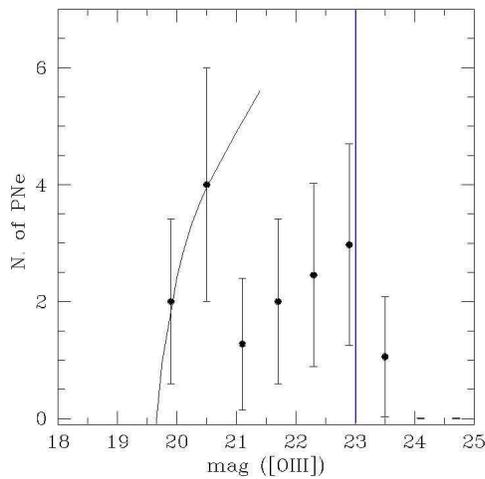

**Fig. 2. PNLF** for the confirmed candidate PNe. The vertical bar represents the completeness limit.

Jacoby & De Marco 2002 with more than 100 PNe in the SMC. They interpret this PNLF dip as arising in young populations in which central star evolution proceeds very quickly; i.e. if many of the PNe were produced during the most recent epoch of star formation. This could also be the case here for NGC 6822.

De Vaucouleur (1978) found a low value $A_B$=0.79 mag for the foreground extinction in the direction of NGC 6822. On the other hand, Lequeux et al. (1980) found $A_V$=1.70 for two H II. Gallart et al. (1996) found a mean value of E(B-V)=0.24±0.03 or $A_V$=0.73. From UBV photometry of OB stars, Massey et al. (1995) found a more complex pattern with the smallest reddening $A_V$=0.79 occurring in the eastern and western sides of the galaxy and the highest values $A_V$=1.37 in the central regions. This would indicate significant internal extinction within NGC 6822, as claimed by Dufour & Talent (1980) who found $A_V$=1.2 for KD_S16. Richer & McCall (1995) found $A_V$=1.28 for KD_S16 and KD_S33.

Correcting for foreground and mean internal reddening (Av~0.8) we derive a true distance modulus of NGC 6822 of m-M=23.3-23.5. The errors are: on photometric zero point (0.15 mag), on the adopted extinction (0.20 mag), on two systematic errors in the definition (0.05 mag) and calibration of the PNLF (0.10 mag error on the Cepheid distance to M 31, Freedman & Madore 1990), and the larger term coming from our best least mean square fit to the PNLF. The total error on the distance modulus is given by the error associated in quadra-ture, whose value should be around 1 mag. Because of the small sample of the 6 PNe used and the complicated Luminosity Function, the statistical significance of this distance is low and much less precise than other previous determinations.

Therefore the PNLF of NGC 6822 with only 6 objects can not be used as a distance indicator. Although it may not be indicative, we note that the distance modulus derived is in agreement with the mean value of 23.48 (van den Bergh 2000) and also with the recent determination of Clementini et al. (2003), based on RR Lyrae (23.36±0.17), or of Cioni (2004), based on the tip of the Red Giant Branch (23.34±0.12).
.

## 6. Perspectives

The new detections of candidate PNe in NGC 6822 and in other nearby galaxies via the Local Group Census have substantially increased the number of known PNe in the LG. Deeper imaging surveys with 8-10m class telescopes would, however, only find fainter PNe. This was shown for instance by very deep imaging with the WFI ESO 2.2m in the LMC and SMC (Leisy et al. 1997), with the WFI and VLT for NGC 3109 (Ismael et al. *in preparation*) or in galaxies beyond the Local Group (NGC 300, NGC 4697, and NGC 1344; Méndez 2004).

The new PNe are clearly an invaluable resource for future spectroscopic studies of individual objects with high S/N spectra, with the aim of at determining their physical and chemical properties and, ultimately, of their host galaxies. This task is devoted to the new 8-10m class telescopes, which can provide high S/N spectra in a reasonable exposure time (2-3 h). This will increase the number of galaxies of the different metallicity, mass, and evolution, for which we can make a direct comparison.

*Acknowledgements.* We are grateful to George Jacoby for sharing his results about PNe positions and the confirmation of most of the objects.

## References

Battinelli, P.; Demers, S.; Letarte, B.; 2003 A&A 405, 563
Bertin, E.; Arnouts, S.; 1996 A&AS 117,393
Clementini, G.; Held, E.V.; Baldacci, L.; Rizzi, L.; 2003 ApJ,588, L85
Chandar, R.; Bianchi, L.; Ford, H.C; 2000 AJ 120, 3088
Ciardullo, R.; Ford, H.C.; Neill, J.D.; Jacoby, G.H.; Shafter, A.W.; 1987 ApJ 318, 520
Ciardullo, R.; Jacoby, G.H.; Ford, H.C.; Neill, J.D.; 1989 ApJ 339, 53
Ciardullo, R.; Jacoby, G.H.; Booth, J.; Ford, H.C.; 1989 ApJ 344, 715
Cioni, MRL.; Habing, HJ.; 2004 Astro-PH 200409294
Collier J.; Hodge P., 1994 ApJS 92, 119
Collier J.; Hodge P.; Kennicutt R.C. Jr; 1995 PASP 107, 361
Corradi, R.L.M.; Magrini L.; Greimel, R.; Irwin, M.; Leisy, P.; Lennon, D.; Mampaso, A.; Perinotto, M., Pollacco, D.R.; Walsh, J.R.; Walton, N.A; Zijlstra, A.A., 2004 A&A submitted
Corradi, R.L.M.; Magrini, L., 2004 ESO Workshop on PNe beyond the Milky Way, Garching, April 2004, in press
de Vaucouleur, G.; 1978 ApJ 223, 730
de Blok, W. J. G.; Walter, F.; 2000 ApJ 537, L95
de Blok, W. J. G.; Walter, F.; 2003 MNRAS 341, L39
D'odorico, S.; Dopita, M. A.; Benvenuti, P.; 1980 A&AS 40, 67
Dufour, R.J.; Talent, D.L.; 1980 ApJ 235, 22
Eddington, A. S.; 1913 MNRAS 73, 359
Ford, H.; Peng, E.; Freeman, K.; 2002 Dynamics, Structure & History of Galaxies, ASP Conf. Pro., Vol. 273, p41
Freedman, W.L. & Madore, B. F.; 1990 ApJ 365, 186
Gallart, C.; Aparicio, A.; Vilchez, J.M.; 1996 AJ 112, 1929
Held, E.V.; Clementini, G.; Rizzi, L.; Momany, Y.; Saviane, I.; Di Fabrizio, L.; 2001 ApJ 562, L39
Hidalgo-Gámez, A.M.; Olofsson, K.; Masegosa, J.; 2001 A&A, 367, 388
Hodge, P.; Kennicutt, R.C. Jr; Lee, M.G.; 1988 PASP 100, 917
Hodge, P.; Lee M.G.; Kennicutt, R.C. Jr; 1989 PASP 101, 32
Hodge, P.; Smith, T.; Eskridge, P.; MacGillivray, H.; Beard, S.; 1991 ApJ 379, 621
Hutchmeier, W.K.; 1979 A&A 75, 170
Irwin, M.; Lewis, J.; 2001 New Astron. 45, 105
Jacoby, G.H.; 1989 ApJ 339, 39
Jacoby, G.H.; 2004 ESO Workshop on PNe beyond the Milky Way, Garching, April 2004, in press
Jacoby, G.H.; De Marco, O.; 2002 AJ 123, 269
Killen, R.M.; Dufour R.J.; 1982 PASP 94, 444
Kinman, T. D.; Green, J. R.; Mahaffey, C. T.; 1979 PASP 91, 749
Leisy, P.; Francois, P.; Fouque, P.; 1997 The Messenger 97, 29
Lequeux, J.; Peimbert, M.; Rayo, J.F.; Serrano, A.; Torres-Peimbert, S.; 1980 A&A 80, 155
Letarte, B.; Demers, S.; Battinelli, P.; Kunkel, W.E.; 2002 AJ 123, 832
Magrini, L.; Corradi, R.L.M.; Mampaso, A.; Perinotto, M.; 2000 A&A, 355, 713
Magrini, L.; Corradi R.L.M.; Walton, N.A.; Zijlstra, A.A.; Pollacco, D.L.; Walsh, J.R.; Perinotto, M.; Lennon, D.J.; Greimel, R.; 2002 A&A, 386, 869
Magrini, L.; Corradi, R.L.M.; Greimel, R.; Leisy, P.; Lennon, D.; Mampaso, A.; Perinotto, M.; Pollacco, D.L.; Walsh, J.R.; Walton, N.A.; Zijlstra, A.A.; 2003 A&A 407, 51
Marconi, G.; Tosi, M.; Greggio, L.; Focardi, P.; 1995 AJ 109, 173
Massey, P.; Armandroff, T.E.; Pyke, R.; Patel, K.; Wilson, C.D.; 1995 AJ 110, 2715
Mateo, M.; 1998 A&A Rev 36, 435
Méndez R.H.; Kudritzki, R.P.; Ciardullo, R.; & Jacoby, G. H.; 1993 A&A 275, 534
Méndez, R.H.; 2004 ESO Workshop on PNe beyond the Milky Way, Garching, April 2004, in press
Monet, D.; Bird, A.; Canzian, B.; et al.; 1998 The USNO-A2.0 Catalog
Muschielok, B.; Kudritzki, R. P.; Appenzeller, I.; Bresolin, F.; Butler, K.; Gässler, W.; Häfner, R.; Hess, H. J.; Hummel, W.; Lennon, D. J.; 1999 A&A 352, L40
O'Dell, C.R.; Hodge, P.W.; Kennicutt, R.C.Jr; 1999 PASP 111, 1382
Richer, M.G.; McCall, M.L.; 1995 ApJ 445, 642
Skillman, E.D.; Evan, D.; Kennicutt, R. C.; Hodge, P.W.; 1989 ApJ 347, 875
van den Bergh, S.; 2000 The galaxies of the Local Group, Cambridge Astrophysics Series, vol no:35
Venn, K.A.; Tolstoy, E.; Kaufer, A.; Skillman, E.D.; Clarkson, S.M.; Smartt, S.J.; Lennon, D.J.; Kudritzki, R.P., 2003 AJ 126, 1326
Wyder, T.K.; 2001 AJ 122, 2490

## Appendix A: Finding charts

We present here the finding-chart of all the candidates. The images shown are taken in the H$\alpha$ filter. North is Up, and East left. The field of view is 1′ 30″.